\documentclass[twocol]{ametsocV6.1}

\usepackage{multirow} 

\title{The ARP-GEM1 Global Atmosphere Model. Part I: Description and Speed-up Analysis}

\authors{David SAINT-MARTIN,\aff{a}\correspondingauthor{David Saint-Martin, david.saint-martin@meteo.fr} and Olivier GEOFFROY\aff{a}}
\affiliation{\aff{a}{CNRM, Université de Toulouse, Météo-France, CNRS, Toulouse, France}}

\abstract{This is the first part of a series of two articles describing the ARP-GEM global atmosphere model version 1 and its evaluation in simulations from 55 km to 6 km resolutions. This article provides a complete description of ARP-GEM1, focusing on its new physical parameterizations and acceleration factors aimed at improving computational efficiency. ARP-GEM1 is approximately 15 times faster than its base model version, ARPEGE-Climat v6.3, while maintaining accurate simulations and enhancing model performance, as shown in Part II.This significant acceleration results from a combination of optimizations, rather than a single factor, with each factor's contribution quantified. Additionally, a detailed decomposition of the model's speedup per component is presented. Streamlining the model reduces computational costs, enhances transparency and portability, and maximizes the benefits of future advanced computing technologies. The results presented here suggest that kilometer-scale global climate simulations should become feasible in the near future.}

\begin{document}

\maketitle

\section{Introduction}

An atmospheric or climate modeling strategy requires compromises between different scientific objectives, where computational cost is a key consideration. The use of computational resources typically follows three main axes, which are often separate and prioritized independently: increasing the number and variety of simulations (e.g., types of simulations, ensemble members, multi-parameter or multi-physics ensembles), enhancing model complexity (by representing additional processes), or increasing model resolution (minimizing unresolved subgrid processes in favor of explicitly resolved ones).

In recent decades, the increase in computational capacity has primarily been absorbed by the increasing complexity of climate Earth System models \citep{schneider-2017}. This complexity arises from the addition of new components and the increased sophistication of existing schemes.
However, the results of this strategy have been disappointing, with CMIP6 models performing similarly to previous generations. Despite reduced errors in large-scale climate metrics due to improved tuning \citep[e.g.,][]{bock-2020, schneider-2024}, the increased complexity in CMIP6 models has not significantly improved climate projections \citep[e.g.,][]{zelinka-2020}. New components like chemistry and biogeochemical cycles do not significantly change the core atmospheric model, and Earth System models often stay close to their parent versions \citep[e.g.,][]{brunner-2020}.

Reducing the computational cost of climate models is increasingly important as global simulations at resolutions that can explicitly resolve deep convection become feasible. Global convection-permitting experiments are currently limited to short durations, typically lasting a few months \citep{stevens-2019}. To enable climate simulations at such resolutions, atmospheric models must become more computationally efficient \citep[e.g.,][]{bauer-2021}.

Reducing the cost of numerical models can be achieved by using next-generation computing platforms, such as Graphics Processing Units. By using these advanced technologies, models can be run more efficiently \citep[e.g.,][]{caldwell-2021}. This makes it possible to perform more complex simulations, run larger model ensembles, or achieve higher resolutions without a proportional increase in computational cost. 

Another strategy involves reducing the model to minimal calculations and essential processes. This approach includes redesigning models to prioritize large-scale advection, necessary radiative computations, and a simplified set of physical parameterizations, ensuring that the model's cost is primarily driven by solving fundamental equations of fluid dynamics \citep{palmer-2019}. Additionally, model speed can be enhanced by optimizing the code through more efficient algorithms (generally reducing memory usage), eliminating redundant calculations, and revisiting numerical methods (e.g., using single precision, spatial and temporal coarsening, or optimizing horizontal and vertical grids). Leveraging new architectures and streamlining calculations are complementary strategies, as improved code efficiency maximizes the benefits of advanced computing technologies.

Streamlining the model not only reduces computational costs but also offers additional benefits: it enhances model transparency and portability to new architectures. Comprehensible physics also favors understanding \citep{held-2005} and, along with the overall simplicity of the model, makes it more accessible to new developers. This strategy, although rarely used or documented, is the approach chosen for the development of the ARP-GEM (Global, Efficient and Multiscale version of ARPEGE) climate model.

The ARP-GEM model version 1 (ARP-GEM1 hereafter) is built upon version 6.3 of ARPEGE‐Climat (v6.3 hereafter), which is described in detail in \citet{roehrig-2020} and earlier technical papers. ARPEGE-Climat v6.3 is the atmospheric component of CNRM-CM6-1 \citep{voldoire-2019} and CNRM-CM6-1-HR \citep{saintmartin-2021}, both of which participated in CMIP6. Like previous versions \citep{deque-1994}, ARPEGE‐Climat is derived from the ARPEGE/IFS (Integrated Forecast System) atmospheric model developed jointly by Météo-France and the European Centre for Medium-Range Weather Forecasts. It is specifically dedicated to climate research applications and may differ from the numerical weather prediction (NWP) version, particularly in its choice of physical parameterizations. Version 6.3 is based on cycle 37 of ARPEGE/IFS. 

The construction of the ARP-GEM1 model involves integrating both original developments and adaptations of recent updates in the ARPEGE/IFS code (from cycle 37 to cycle 46). The main developments are described in the following sections. Compared to v6.3, the ARP-GEM1 model benefits from: changes in the dynamics (Section \ref{sec:dynamics}), a revised and significantly improved suite of physical parameterizations (Section \ref{sec:physics}), and a set of developments aimed at improving code efficiency, referred to as 'acceleration factors' (detailed in Section \ref{sec:accelerations}). Section \ref{sec:results} presents an analysis of the model’s performance results and discusses their role in the model’s overall acceleration. An assessment of the model’s accuracy in representing climate across 30-year prescribed sea-surface-temperature simulations at 55, 25, 12, and 6 km resolutions is presented in Part II of this study (Geoffroy and Saint-Martin 2024, hereafter Part II).

\section{Dynamical core}
\label{sec:dynamics}

\subsection{Overview}
\label{sec:dyn_overview}
Atmospheric models approximate atmospheric flow by numerically solving a set of nonlinear partial differential equations, known as the 'primitive' equations. In the ARP-GEM1 model, the primitive equations follow the standard hydrostatic assumption. Like most operational NWP models, the dynamical core uses the efficient semi-implicit and semi-Lagrangian discretization methods to solve these equations. The prognostic variables include winds, surface pressure, temperature, specific humidity and the specific water contents of four hydrometeor types: cloud liquid, cloud ice, rain, and snow.

\subsection{The semi-Lagrangian scheme}
\label{sec:SL_scheme}
The semi-Lagrangian (SL) advection scheme evaluates the transport of a quantity by tracing back from a grid point to the departure point, which is the location where a parcel of air originated at the previous timestep. \citep[e.g.,][]{staniforth-1991}. The advected quantity is equal to the interpolated value of the quantity at the departure point, using values from surrounding grid points. This method is termed semi-Lagrangian because the parcel is constrained to arrive at a grid point by the end of the timestep. The choice of the interpolation method is a key element of the SL scheme.

Although the SL scheme is accurate and efficient, it is not conservative \citep[e.g.,][]{williamson-1994}. Without sources or sinks, the global mass of a tracer may not remain constant, especially for tracers with sharp gradients. The use of high-order interpolation methods to advect cloud liquid, cloud ice, rain, and snow (as in v6.3) can result in spurious extrema and negative values, leading to artificial increases in water mass when clipped. To counteract this deficiency, ARPEGE-Climat v6.3 applies a local mass fixer algorithm to the specific humidity and hydrometeors, adjusting the solution mainly in regions where gradients are large \citep{diamantakis-2014}. In the ARP-GEM1 model, we choose to address this deficiency by replacing the cubic horizontal interpolation with a linear horizontal interpolation scheme for the four prognostic hydrometeors, as also reported in \citet{rackow-2024}. This change significantly reduces water mass non-conservation and allows the mass fixer to be disabled (see also Section \ref{sec:accelerations}\ref{sec:conserv}).

For temperature and humidity, we also replace the quasi-cubic interpolation method by a quintic vertical interpolation \citep{polichtchouk-2020}, which means that a Lagrange polynomial of degree 5 (instead of 3) interpolates a field using 6 neighbouring points (instead of 4). \citet{polichtchouk-2020} shows that increasing the order of vertical interpolation can serve as an alternative to adding more vertical levels.

\begin{table*}
\caption{Moist physical parameterizations of v6.3, v6.5 and ARP-GEM1.}
\label{tab:phys}
\begin{center}
\begin{tabular}{llll}
\topline
& ARPEGE v6.3 & ARPEGE v6.5 & ARP-GEM1 \\
\midline
Large-scale clouds & \citet{bougeault-1982} & \citet{smith-1990} & \citet{smith-1990} \\
Turbulence & CBR00 & CBR00 & Modified CBR00 \\
\ \ \ \ \ Mixing length & BL89 & BL89 & Modified BL89 \\
\ \ \ \ \ Moist processes & \citet{bougeault-1982} & \citet{bougeault-1982} & \citet{bretherton-2009} \\
Shallow convection & \multirow{2}{*}{\citet{piriou-2007, gueremy-2011}} & Modified PMMC09 & MFUP \\
Deep convection &  & Tiedtke-Bechtold  & Modified Tiedtke-Bechtold \\
Microphysics & \citet{lopez-2002}& \citet{lopez-2002} & Modified \citet{lopez-2002} \\
\botline
\end{tabular}
\end{center}
\end{table*}

\subsection{The semi-implicit scheme}
In ARP-GEM1, we use the two-time-level and time-centered form of the semi-implicit (SI) time integration method, which it is nearly equivalent to the Crank-Nicholson scheme \citep[e.g.,][]{temperton-1987}. The semi-implicit time discretisation is derived by subtracting from the primitive equations a system of equations linearised around a reference state. The linear part is treated implicitly, whereas the discretisation of the nonlinear residual is explicit \citep{benard-2003}. The resulting system of equations can be reduced to a single Helmholtz equation.

\subsection{The spectral method}
This linear system is solved efficiently by partitioning the computations between physical (grid-point) space and spectral (spherical-harmonics) space. Semi-Lagrangian advection and physical parameterizations are performed in grid-point space, while the Helmholtz equation and horizontal diffusion are handled in spectral space. This partitioning involves spectral transforms, combining Legendre and Fourier transforms to convert fields between spectral and grid-point spaces. 

In the ARP-GEM1 model, the use of a cubic truncation spherical-harmonics grid (see Section \ref{sec:accelerations}\ref{sec:cubic}) enhances the efficiency of the spectral method without sacrificing accuracy \citep{wedi-2014}. This cubic truncation avoids aliasing and acts as a numerical filter, requiring minimal numerical diffusion. The horizontal numerical diffusion scheme used in ARPEGE has thus been replaced by a simple scheme based on spectral viscosity \citep{gelb-2001}, resulting in computational gains. 

Another important change (and computational gain) in the spectral method concerns specific humidity. We choose not to apply numerical diffusion to specific humidity. Thus, the transformation from spectral space to grid-point space becomes unnecessary for specific humidity, allowing us to save computational cost on the spectral transformation of a 3D field. The discrete Fourier transform is performed with the Fast Fourier Transform (FFT) package FFTW\footnote{http://www.fftw.org}, which efficiently allows any number of points per latitude circle, implying the use of an octahedral physical-space grid. For the Legendre Transform, the Fast Legendre Transform method \citep[FLT,][]{wedi-2013} is used for the high-resolution ($\leq$ 12 km) versions of the ARP-GEM1 model (see Part II).

\subsection{Stability and timestep}
The value of the timestep plays a major role in the efficiency of an atmospheric model. Each of these numerical methods (SI, SL) relaxes computational stability constraints and allows the use of large timesteps. The stability condition of the SL scheme is less severe than the CFL condition and the SI scheme removes fast but energetically insignificant waves. In the ARP-GEM1 model, the choice of the dynamical core plus a work for stabilizing physical schemes allows the use of a timestep of 600 s at 12 km and 300 s at 6 km. At 55 km, we use a default timestep of 900 s. Nevertheless, long-term simulations remain stable with a timestep of 1200 s. 

\section{Physical parameterizations}
\label{sec:physics}

\subsection{Overview}
\label{physics:overview}
The physical suite of parameterizations in ARPEGE-Climat v6.3 has been significantly revised in ARPEGE-Climat version 6.5 (hereafter v6.5). In v6.5, we added a shallow convection scheme and replaced the large-scale cloud scheme, the convection scheme, and the radiation scheme. 

The v6.5 large-scale cloud scheme is the \citet{smith-1990} scheme, which is employed in the NWP version of ARPEGE. The deep convection scheme is originally described in \citet{tiedtke-1989} and revised by \citet{bechtold-2008, bechtold-2014}. It will be referred to as the Tiedtke-Bechtold scheme, recently implemented in NWP version. The radiation scheme (ecRad) is described in \citet{hogan-2016}. Lastly, v6.5 incorporates a modified shallow convection scheme from the Meso-NH model \citep{lac-2018} based on \citet[][hereafter PMMC09]{pergaud-2009} with entrainment formulations from \citet{rio-2010}. Additionally, it includes developments such as shallow precipitation, detrainment, and revised numerics.

The physics used in the ARP-GEM1 model is an improved and updated version of that in v6.5. The main physical packages used in v6.3, v6.5, and ARP-GEM1 are summarized in Table \ref{tab:phys}. The new physics in ARP-GEM1 is described in detail in the following sections.

\subsection{Large scale cloud}
\label{physics:cloud}
The v6.3 large-scale cloud scheme is based on the statistical cloud scheme of \citet{sommeria-1977} and the \citet{mellor-1977} scheme, revisited by \citet{bougeault-1982}. Bougeault’s approach uses an asymmetric distribution (gaussian-exponential) to represent the subgrid variability of total humidity and liquid water temperature $T_{l}$. The \citet{mellor-1977} scheme employs a Gaussian probability density function (PDF).

In these schemes, the standard deviation of the PDF is related to turbulent mixing using the same formalism as other turbulent quantities. This approach seems physically reasonable in the boundary layer (disregarding interaction with shallow convection), but it may be limited at upper levels where distributions are large \citep[e.g.,][]{quaas-2012} while TKE is small \citep[e.g.,][]{de_rooy-2022}.

In ARP-GEM1, we use the \citet{smith-1990} cloud scheme for its simplicity. The standard deviation of the subgrid PDF of the difference between subgrid specific humidity and subgrid saturation specific humidity is proportional to the grid-point mean saturation specific humidity associated with the liquid water temperature. The scaling coefficient depends on the critical relative humidity $RH_\mathrm{c}$, which is the subgrid relative humidity threshold above which condensation occurs.

The $RH_\mathrm{c}$ parameter is commonly set as a function of height in climate models. In v6.5, the $RH_\mathrm{c}$ profile is a third-order polynomial that decreases from the surface to a fixed level corresponding to an altitude of about 400 hPa and remains constant above. In the present version, the $RH_\mathrm{c}$ profile is similar to that used in CAM4, with a constant value at low levels ($RH_{\mathrm{c},\mathrm{low}} = 0.94$), a constant value at high levels ($RH_{\mathrm{c},\mathrm{high}} = 0.70$), and linear interpolation in between. The layer limits are specified as parameters, which can be adjusted for model tuning if needed, and are set to 700 hPa and 500 hPa. The values of $RH_\mathrm{c}$, determined through tuning, are consistent with \citet{quaas-2012}.

In the cloud scheme, updated values of $T$, $q_v$, $q_l$, and $q_i$ are computed. These values are then used throughout the code to compute other physical processes. In v6.3 and v6.5, some variables used in subsequent routines were updated in the cloud scheme, while others were carried over from previous timesteps. If not impacting model results, this issue has been addressed in ARP-GEM1 to ensure consistency throughout the model and favor readability.

\subsection{Turbulence}
\label{physics:turbulence}
In all model versions, the turbulence scheme is from \cite{cuxart-2000} scheme (hereafter CBR00) except in upper atmosphere where it follows the simple parameterization of \citet{louis-1979}. Note that TKE is a semi-prognostic variable, i.e., it is stored at each timestep but not advected.

\subsubsection{Buoyancy production}
In the CBR00 scheme, the subgrid PDF of humidity and temperature must be known to compute the buoyancy flux (then the thermal buoyancy production of the TKE), which necessitate the knowledge of cloud variables. In v6.3 and v6.5, the CBR00 turbulence scheme computes its own large-scale clouds based on the \citet{bougeault-1982} distributions. This lack of consistency when using the \citet{smith-1990} scheme results in a redundant calculation.
In ARP-GEM1, the expression of buoyancy flux is simplified following \citet{bretherton-2009}:
\begin{equation}
\overline{w'\theta_v'} = k_{\theta} \overline{w'\theta_l'} + k_{q}T \overline{w'q_t'}
\end{equation}
with
\begin{equation}
k_{\theta} = (1-C^{\mathrm{ls}}_{k-1/2}) a_{d} + C^{\mathrm{ls}}_{k-1/2} a_{m}
\end{equation}
\begin{equation}
k_{q} = (1-C^{\mathrm{ls}}_{k-1/2}) b_{d} + C^{\mathrm{ls}}_{k-1/2} b_{m} \overline{w'q_t'}
\end{equation}
Large-scale cloud fractions at interface layers, $\{C^{\mathrm{ls}}_{k-1/2}\}_{k \in \{1,..,K+1\}}$, depend on values at the $K$ 'full' layers, $\{C^{\mathrm{ls}}_{k}\}_{k \in \{1,..,K\}}$:
\begin{equation}
C^{\mathrm{ls}}_{k-1/2} = \min \left(C^{\mathrm{ls}}_{k-1}, 0.5(C^{\mathrm{ls}}_{k} + C^{\mathrm{ls}}_{k-1})\right)
\end{equation}
$a_{d}$, $b_{d}$, $a_{m}$ and $b_{m}$ are dry and moist thermodynamic coefficients relating thermodynamic variables involved in unsaturated and saturated conditions \citep{stevens-2002}, with following values: $a_{d}=1$, $b_{d}=R_v/R_d-1 (\approx 0.608$), $a_{m}=0.5$, $b_{m}=3.5$. In the present version, liquid values are used even for ice conditions. Note that during the stage of model development, this simplification was not found to impact the model results.

\subsubsection{Moist mixing length}
In v6.3, turbulent mixing length is computed from upward and downward mixing lengths ($L_{\mathrm{up}}$ and $L_{\mathrm{dw}}$) that are derived from adiabatic upward and downward vertical displacements of an undiluted parcel with initial kinetic energy equal to the environmental TKE \citep[][hereafter BL89]{bougeault-1989}. The scheme assumes dry conditions. 

In ARP-GEM1, we take into account moist processes. The degree of saturation of the displaced parcel is assumed to be that of the environment as in \citet{saanchez-2004}. By using the same formalism to represent moist processes in the buoyancy flux computation, the upward and downward mixing length equation reads:
\begin{equation}
e(z) = \int_z^{z+L_{\mathrm{up}}} g \left[1- \theta_{v,p}^\uparrow (z,z') / \theta_{v,e}^\uparrow(z') \right] dz'
\end{equation}
\begin{equation}
e(z) = \int_{z-L_{\mathrm{dw}}}^{z} g \left[1- \theta_{v,p}^\downarrow (z,z') / \theta_{v,e}^\downarrow (z') \right] dz'
\end{equation}
with linearized virtual potential temperature \citep[e.g.,][]{bellon-2016} of the displaced parcel $\theta_{v,p}$ and that of the environment $\theta_{v,e}$:

\begin{eqnarray}
\theta_{v,p}^\uparrow (z,z_k) & = & \left[(C_{k}^\uparrow-1) a_d + C_{k}^\uparrow a_m\right] \theta_{l}(z) \\ \nonumber
 & + & \left[(C_{k}^\uparrow-1) b_d + C_{k}^\uparrow b_m\right] \theta_{k} q_{t}(z) 
\end{eqnarray}

\begin{eqnarray}
\theta_{v,e}^\uparrow(z_k) & = & \left[(C_{k}^\uparrow-1) a_d + C_{k}^\uparrow a_m\right] \theta_{l}(z_k) \\ \nonumber
 & + & \left[(C_{k}^\uparrow-1) b_d + C_{k}^\uparrow b_m\right] \theta_{k} q_{t}(z_k) 
\end{eqnarray}
and similar expression for downward displacements; $\theta_{k}$ is potential temperature at level k, $C_{k}^\uparrow = C_{k}^{\mathrm{ls}}$ and $C_{k}^\downarrow = \min ( C_{k}^{\mathrm{ls}},C_{k-1}^{\mathrm{ls}} )$.
The inclusion of moist processes leads to increased vertical mixing in cloudy boundary layers. 

In v6.3, the final computation of the mixing length $L_m$ follows Eq. 12a in \citet{lemarie-2021}. In ARP-GEM1, a simpler formulation is used (with the exponents $\alpha$ set to 1, instead of 2/3 in v6.3) and with the introduction of a scaling coefficient $\mathcal{C}_L$ depending on turbulence parameters such that it satisfies the limit at ground presented in \citet{lemarie-2021} and \citet{lenderink-2004}:
\begin{equation}
\frac{1}{L_m} = \mathcal{C}_L \left( \frac{1}{L_{\mathrm{up}}} + \frac{1}{L_{\mathrm{dw}}} \right)
\end{equation}
with $\mathcal{C}_L$ = 0.51, close to a harmonic average.

\subsubsection{Cloudy boundary-layer top entrainment}
A boundary layer cloud-top  entrainment is parameterized following the simple formulation used in IFS \citep{ecmwf-2019}:
\begin{equation}
K_{M}= \max(K_M, 0.75 K_{e})
\end{equation}
\begin{equation}K_{H}= \max(K_H, K_{e})
\end{equation}
$K_M$ and $K_H$ are eddy diffusivity coefficients for momentum and thermodynamic variables, respectively, and $K_e$ is a entrainment diffusivity coefficient:
\begin{equation}
K_{e} = -0.2 \frac{\Delta R_{\mathrm{lw}}}{\rho c_{p}} \frac{\Delta z}{\Delta \theta_v}
\end{equation}
where $\Delta R_{\mathrm{lw}}$ is the longwave radiative flux jump at cloud top, and $\Delta\theta_v$ is the jump in virtual potential temperature. In IFS, the formulation is applied only in case of stratocumulus boundary layer, as defined from estimation inversion strength criteria. In ARP-GEM1, this formulation is applied at top of shallow convective updraft and if a large-scale cloud fraction is present.

\subsection{Dry and moist shallow convection}
\label{physics:shallow}
The new shallow convection scheme is a mass-flux scheme that describes a single bulk thermal plume model, representing both dry and moist thermals \citep[e.g.,][]{pergaud-2009}. It is referred to as MFUP (Mass Flux UPward). 
It has been developed from the routine that describes the mass-flux component of the IFS boundary-layer scheme \citep{siebesma-1995, ecmwf-2019}. In IFS, this scheme is separate from the shallow convection scheme and includes simple physics. We use this scheme as the foundation for MFUP, with extensive revisions including vertical velocity, lateral entrainment and detrainment, convective closure, discretizations, the addition of limiting mixing at strong inversions, and the incorporation of diagnostic shallow clouds and precipitation. 

\subsubsection{Updraft initialisations}
\label{sec:updraftinit}
Updrafts originate from the first atmospheric model interface (the interface between the first and second model layers, $K-1/2$). Following the approach of the original IFS routine and the Tiedtke-Bechtold scheme, the initial updraft velocity ($w_u$), static energy excess ($\Delta s_u$), and humidity excess ($\Delta q_{t,u}$) are assumed to scale with surface fluxes \citep{ecmwf-2019}. We use formulations from \citet{lenschow-1980} rather than those of \citet{jakob-2003}:
\begin{equation}
w_u(z_{K-1/2}) = g\frac{\overline{w'\theta_v'}|_{\mathrm{surf}}}{\theta} (z_{K-1/2})^{1/3}
\label{eq:wini}
\end{equation}
\begin{equation}
\Delta s_u(z_{K-1/2}) = \mathcal{C}_{\mathrm{ini}} \frac{\overline{w's'}|_{\mathrm{surf}}} { w_u(z_{K-1/2})}
\label{eq:sini}
\end{equation}
\begin{equation}
\Delta q_{t,u}(z_{K-1/2}) = \mathcal{C}_{\mathrm{ini}} \frac{\overline{w'q_t'}|_{\mathrm{surf}}} { w_u(z_{K-1/2})}
\label{eq:qini}
\end{equation}
The proportionality constant $\mathcal{C}_{\mathrm{ini}}$ is set to 1.2, rather than 1.5 in \citet{lenschow-1980}.

\subsubsection{Updraft vertical velocity}
The parcel updraft vertical velocity equation \citep[e.g.,][]{de_roode-2012} is:
\begin{equation}
\frac{1}{2}\frac{\partial w_u^2}{\partial z} = a B_u - d w_u^2
\label{eq:vver}
\end{equation}
where $B_u$ is the parcel buoyancy, $d$ is a drag coefficient set to a constant value ($d = 0.0013$ m$^{-1}$) as in \citet{rio-2010}. $a$ is equal to $2/3$ for $B_u<$ 0 and is reduced to $1/3$ for $B_u>0$, constituting a contribution to the drag. This aim to represent effects of pressure perturbation \citep[e.g.,][]{de_roode-2012}.

\subsubsection{Lateral entrainment}
In dry conditions, the narrowing of the area fraction associated with updraft acceleration is assumed to be roughly compensated by entrainment \citep{nordeng-1994,rio-2010}. The fractional entrainment of dry updraft is given by:
\begin{equation}
\epsilon_d = \max \left(\frac{1}{w}\frac{\delta w}{\delta z}, \epsilon_{d0} \right)
\end{equation}
with $\epsilon_{d0} = $ 0.0001 m$^{-1}$.
In saturated updrafts, the fractional entrainment rate follows a $B/w^2$ dependency \citep{fox-1970,gregory-2001,del_genio-2010}:
\begin{equation}
\epsilon_m = \max \left(0.25 \frac{B}{w^2}\ ,  \epsilon_{m0} \right)
\end{equation}
with $\epsilon_{m0} = $ 0.0005 m$^{-1}$. 
When parameterizing entrainment based on the buoyancy sorting mechanism \citep{kain-1990,bretherton-2004}, the entrainment rate increases with relative humidity. However, some studies suggest that it has to decrease with relative humidity \citep[e.g.,][]{bechtold-2008}. Additionally, the validity of the buoyancy sorting mechanism for representing entrainment is unclear \citep[e.g.,][]{de_rooy-2013}. Therefore, we do not include a dependency on humidity in our entrainment formulation, in contrast to the detrainment. 

\subsubsection{Lateral detrainment}
The fractional detrainment rate, $\delta$, is parameterized as follows: 
\begin{equation}
\delta = \delta_{w} + \alpha_{\mathrm{dry}}\delta_{d}+(1-\alpha_{\mathrm{dry}})\delta_{m}
\label{eq:detrainment}
\end{equation}
The first term is a dynamical term related to the parcel’s deceleration \citep{rio-2010}:
\begin{equation}
\delta_{w} = \max(-\beta_{\delta}\frac{1}{w}\frac{\delta w}{\delta z},0)
\end{equation}
 with $\beta_{\delta}$ equal to 1.3.
The second and third terms represent additional entrainment terms for dry and moist cases, respectively. The coefficient $\alpha_{\mathrm{dry}}$ denotes the fraction of dry thermals in the considered layer. It is equal to one below the cloud base level ($z_{b}$), zero above the cloud base level, and $(z_{b} - z_{k-1/2}) / (z_{k+1/2}-z_{k-1/2})$ at the cloud base level, as computed in the original IFS routine.

The dry fractional detrainment $\delta_{d}$ is similar to a turbulent detrainment term \cite[e.g.,][]{de_rooy-2013} and is simply set to a constant value (equal to 0.00065 m$^{-1}$). 

The third term of Eq. \eqref{eq:detrainment} is related to the mixing of cloudy air with the environmental dry air, which causes buoyancy reversal and detrainment. It may primarily represent the detrainment of the mean cloud population as a whole (considering clouds have varying depths) rather than that of each individual cloud. It is parameterized through the buoyancy sorting approach \citep{kain-1990,zhao-2003, bretherton-2004, de_rooy-2008, boing-2012} with a bound set to $\delta_{d}$, ensuring that it is non-zero in saturated environments (hence stratocumulus):
\begin{equation}
\delta_{m}= \max(\delta_0 (1-\chi_c)^2, \delta_{d})
\end{equation}
with $\delta_0 = 0.006$.
The variable $\chi_c$ is the critical mixing fraction of the environmental air above which a mixture of ascending air and environmental air becomes negatively buoyant with respect to the environment \citep{kain-1990}. It is calculated following the formula of \citet{de_rooy-2008}:
\begin{equation}
\chi_c = \frac{ c_p \Delta T_{v} }{ \beta c_p \Delta T_{l} + (\beta-\alpha) L \Delta q_{t}}
\end{equation}
with $\alpha=0.12$ and $\beta=0.4$, and bounds set to 0 and 1. $\Delta T_{v} = T_v^{\mathrm{up}}-T_v^{\mathrm{e}}$, $\Delta T_{l} = T_l^{\mathrm{up}}-T_l^{\mathrm{e}}$, and $\Delta q_{t} = q_t^{\mathrm{up}}-q_t^{\mathrm{e}}$ denote the updraft excesses. Note that liquid cloud values are also used for ice conditions.

\subsubsection{Limitation of the cloud top mixing}
Above the level of neutral buoyancy, negatively buoyant mixtures can sink back below the inversion \citep[e.g.,][]{wyant-1997, bretherton-1997,hourdin-2019}. If mass flux is too strong at the inversion level, the enhanced mixing at cloud top can break down stratocumulus decks.

In the IFS model, when a stratocumulus is diagnosed (based on an estimated inversion strength criteria), shallow convection is not activated and boundary-layer mass flux at the inversion is set to 0. In the scheme of \citet{bretherton-2004}, all negatively buoyant mixtures in the overshooting zone (above the inversion) are detrained  below it. In the scheme of \citet{hourdin-2019}, the parcel buoyancy entering the vertical velocity equation is computed by using the environmental potential temperature at a higher level through the use of a parameter. In v6.5, environmental properties are averaged between two vertical level interfaces. This approach has a similar effect to the method of \citet{hourdin-2019} but does not introduce a parameter that would require calibration and may depend on vertical resolution.

In ARP-GEM1, we apply a limitation of the amount of air mass overshooted at the inversion level. At the altitude $\Delta z_{\mathrm{o}}$ above the highest interfacial layer with positive velocity (denoted by $k_{\mathrm{top}} - 1/2$), the updraft velocity, $w_u(z_{k_{\mathrm{top}} - 1/2} + \Delta z_{\mathrm{o}})$, is computed with Eq. \eqref{eq:vver}. If this velocity is zero, all the air is assumed to detrain in the underlying level $k_{\mathrm{top}}+1$ and the mass flux at altitude $z_{k_{\mathrm{top}} - 1/2}$ is set to 0. If the velocity is positive, the mass flux is simply rescaled by $w_u^2(z_{k_{\mathrm{top}} - 1/2}+\Delta z_{\mathrm{o}}) / w_u^2(z_{k_{\mathrm{top}} - 1/2})$. The depth $\Delta z_{\mathrm{o}}$ is a parameter, set to 25 m. The amount of low-level clouds is sensitive to this parameter.

Finally, there is redundancy between the use of parameterized cloud top entrainment (see Section \ref{sec:physics}\ref{physics:turbulence}) and shallow mixing at the boundary-layer top inversion. This lack of coherence between the turbulence and shallow convection schemes will be addressed in future versions.

\subsubsection{Convective closure}
We use a surface flux closure formulation  where the mass flux is assumed to scale with the convective scale velocity $w_*$ \citep{grant-2001, pergaud-2009}. The equality is set at the first model interface rather than at the surface \citep{pergaud-2009} or at cloud base 
\citep{grant-2001}. Given the formulation of the initial vertical velocity (Eq. \eqref{eq:wini}), this closure is roughly equivalent to an initial convective area fraction $\alpha_{u}$ scaling with $z_{i}^{1/3}$: 
\begin{equation}
\alpha_{u}(z_{K-1/2}) = \min(\mathcal{C}_M z_{i}^{1/3},\alpha_{u,\text{max}})
\end{equation}
with $\mathcal{C}_M=0.015$ m$^{-1/3}$ and $\alpha_{u,\text{max}}=0.30$ in this model version. In \cite{pergaud-2009}, the boundary layer depth $z_{i}$ is computed from the \citet{bougeault-1989} upward mixing length at surface. Here, it is calculated as the altitude where a dry updraft buoyancy becomes negative. This updraft is initialized with the initial properties described in Section \ref{sec:physics}\ref{physics:shallow}.\ref{sec:updraftinit}, and its fractional entrainment rate equates $0.7/z$.  

\subsubsection{Shallow convective clouds and precipitation}
Shallow convective cloud properties are simply derived from the convective area fraction and the updraft specific water content ($q_{c}^{\mathrm{up}}=q_{l}^{\mathrm{up}}+q_{i}^{\mathrm{up}}$), both defined at the interface layers:
\begin{equation}
C^{\mathrm{sh}}_k = k_{\mathrm{cld}}\left(\alpha_{u,k-1/2}+\alpha_{u,k+1/2}\right)/2
\end{equation}
\begin{equation}
q^{\mathrm{sh}}_{c,k} = C^{\mathrm{sh}}_k \left(q^{\mathrm{up}}_{c, k-1/2}+q^{\mathrm{up}}_{c, k+1/2} \right)/2
\end{equation}
with $k_\mathrm{cld}$ set to a high value of 2.4 in this model version.
The liquid and ice specific water contents are derived from the specific cloud water content using the same cloud-ice partitioning as for large-scale clouds.

The shallow convective precipitation scheme is a diagnostic scheme \citep{sundqvist-1988, neale-2013}. The precipitation rate $P_{r}$ at a given interfacial layer is given by:
\begin{equation}
\mathcal{P}^{\mathrm{sh}}_{k+1/2} = \mathcal{P}^{\mathrm{sh}}_{k-1/2} + \alpha_u \left[A^{\mathrm{sh}}_k + E^{\mathrm{sh}}_k\right] \frac{\Delta p}{g}
\end{equation}
where $\Delta p/g$ is the depth of the layer. Precipitation is defined as rain if environmental temperature is greater than 0$^\circ$C and snow elsewhere.

The autoconversion rate $A^{\mathrm{sh}}$ is parameterized by using Kessler-type formulation:
\begin{equation}
A^{\mathrm{sh}} = k_{\mathrm{au},l}^{\mathrm{sh}} \max(0, q_{l}^{\mathrm{up}} - q_{l0}^{\mathrm{sh}} ) + k_{\mathrm{au},i}^{\mathrm{sh}} \max(0, q_{i}^{\mathrm{up}} - q_{i0}^{\mathrm{sh}} )
\end{equation}
with a different critical liquid water over land and ocean:
\begin{eqnarray}
q_{l0}^{\mathrm{sh}} = (1-\alpha^{\text{land}})q_{l0}^{\text{ocean}} + \alpha^{\text{land}} q_{l0}^{\text{land}}
\end{eqnarray}
with $k_{\mathrm{au},l}^{\mathrm{sh}} = k_{\mathrm{au},i}^{\mathrm{sh}} = 3 \cdot 10^{-3}$ s$^{-1}$, $q_{l0}^{\text{ocean}} = 300 \cdot 10^{-6}$ kg.kg$^{-1}$, $q_{l0}^{\text{land}}=800 \cdot 10^{-6}$ kg.kg$^{-1}$, and $q_{i0}^{\mathrm{sh}}=20 \cdot 10^{-6}$ kg.kg$^{-1}$.  The evaporation rate $E$ is parameterized following \cite{sundqvist-1988}: 
\begin{equation}
E_k^{\mathrm{sh}} = - 1.5 \cdot 10^{-5} \max(0, 1 - RH) (\mathcal{P}^{\mathrm{sh}}_{k-1/2})^{1/2}
\end{equation}

\subsection{Deep convection}
\label{physics:deep}
In ARP-GEM1, the deep convection scheme is mainly based on the Tiedtke-Bechtold scheme. A full description of this scheme can be found in \citet[][and references therein]{ecmwf-2019}. 
The Tiedtke-Bechtold scheme incorporates an updated version of the \citet{jakob-2003} triggering method. In v6.5, the initial test parcel for deep convection starts from the second model level interface. If the criterion for a deep updraft is not met at this level, the next level is tested, and so on, until a maximum starting altitude of 250 hPa is reached. When starting from the lowest 60 hPa of the atmosphere, the thermodynamic properties of the test parcel are averaged from the layer in question and its two adjacent layers.

In ARP-GEM1, the number of potential start levels is reduced, thereby decreasing the numerical cost of triggering. Initially, the test parcel for deep convection starts from the top of the 40 hPa depth layer, using the mean thermodynamic properties of all levels within this layer, similarly to \citet{kain-2004}. The maximum starting altitude is then set to correspond to the top of the boundary layer, fixed at an altitude of approximately 750 hPa (instead of 250 hPa), which is consistent with other parameterizations \citep[][and references therein]{villalba-pradas-2022}. These modifications were not found to significantly alter precipitation patterns during model development.

In addition, we have removed all computations related to shallow convection. In v6.5, most calculations are performed with all final tendencies set to zero. Additionally, the parameterization for mid-level convection has been removed, as many models do not use such a parameterization.

We use the parameter values of \citet{ecmwf-2019}, except for four exceptions. The deep convective minimum cloud depth allowed -- the threshold used to define deep convection -- is set to 300 hPa (instead of 200 hPa). Note that the original \citet{tiedtke-1989} scheme used this value. The autoconversion rate over ocean $c_{00}$ is fixed at 1.35$\cdot$10$^{-3}$ s$^{-1}$ and the entrainment rate coefficient $\epsilon_{\text{up}}$ is set to 1.8$\cdot$10$^{-3}$ s$^{-3}$.m$^{-1}$. Finally, the convective closure time scale is rescaled by a parameter, $k_{\mathrm{cv}}$ (=$1/\alpha_x$ in \citet{ecmwf-2019}). The parameters $k_{\mathrm{cv}}$ is set to a constant value equal to $1.35$ in 55-km simulations. This parameters can be allowed to vary with resolution. (\cite{ecmwf-2019}).

\subsection{Large-scale microphysics}
\label{physics:micro}
The large-scale microphysics scheme is based on \citet{lopez-2002}, combined with the cost-efficient sedimentation scheme of \cite{bouteloup-2011}, as used in v6.3 and in v6.5. It represents two precipitation species - rain ($q_{r}$) and snow ($q_{s}$) - and two cloud species - liquid ($q_{l}$) and ice ($q_{i}$). This minimal set of precipitating hydrometeors, along with simple process representations (e.g., Kessler-type autoconversions, constant vertical velocities, simple evaporation, and no representation of rain overlap), classifies it as a minimal bulk microphysics scheme \citep{yano-2010}. Given the complexity of microphysics, such a minimal scheme may introduce only minor additional errors compared to more intricate schemes, while enhancing ease of understanding and tuning.

Minor differences with the original version are introduced in ARP-GEM1. The ice concentration (used for sublimation) is not an exponential function of temperature anymore (Appendix D in \citet{lopez-2002}). The intercept parameter in snow size distribution $N_{0s}$ is set to a constant value equal to $12 \cdot 10^6$ m$^{-4}$. The intercept parameter in liquid particle size distribution $N_{0r}$ is set to $4 \cdot 10^6$ m$^{-4}$. These values are consistent with \citet{lin-1983}.
Constant sedimentation velocities are set to 4 m.s$^{-1}$ for rain and 0.9 m.s$^{-1}$ for snow and 0.06 m.s$^{-1}$ for ice. The cloud droplet sedimentation process is removed. 

The evaporation and sublimation formulations are simplified:
\begin{equation}
E_{r} = \mathcal{C}_{r} \frac{N_{0r}(1-C^{\mathrm{ls}}) (1-q_{l}/q_{\mathrm{sat},l})}{\rho( K_r + D_r )}{q_r}^{0.5}
\end{equation}
\begin{equation}
S_{s} = \mathcal{C}_{s} \frac{N_{0s}(1-C^{\mathrm{ls}}) (1-q_{i}/q_{\mathrm{sat},i})}{\rho( K_s + D_s )}{q_s}^{0.5}
\end{equation}
\begin{equation}
K_{r/s}\approx \frac{L_{v/s}^2}{K_tR_vT^2}
\end{equation}
\begin{equation}
D_{r/s}\approx \frac{R_dT}{2 q_{\mathrm{sat},l/i}}
\end{equation}
$q_{\mathrm{sat},l}$ and $q_{\mathrm{sat},i}$ are saturation specific humidity with respect to liquid and ice, $L_{v}$ and $L_{s}$ are latent heats of vaporization and sublimation, $R_v$ is specific gas constant of water vapor, $K_t$ is thermal conductivity of air, $R_d$ is specific gas constant of dry air. Values are given in \cite{lopez-2002}.
Finally, $\mathcal{C}_{r}= 8 \cdot 10^{-5}$ SI and $\mathcal{C}_{s} = 4.7 \cdot 10^{-5}$ SI.

As in \citet{lopez-2002}, rain and snow autoconversion, $A_r$ and $A_s$, follows the \citet{lin-1983} formulations:
\begin{equation}
A_r = k_{\mathrm{au},l}C^{\mathrm{ls}} \max( 0, q_l/C^{\mathrm{ls}} - q_{l0} )
\end{equation}
\begin{equation}
A_s = k_{\mathrm{au},i}C^{\mathrm{ls}} \max( 0, q_i/C^{\mathrm{ls}} - q_{i0}(T))
\end{equation}

The value of the ice autoconversion coefficient $k_{\mathrm{au},i}$ was  widely revised in the tuning process in comparison with v6.3. It was reduced of more than a factor 5, in association with improvement in the representation of high clouds. At 55 km, a value of $7\cdot10^{-4} s^{-1}$ for $k_{\mathrm{au},i}$ is used. This order of magnitude is consistent with values of \cite{lin-1983}, and those used in CAM4 \citep{neale-2013} or IFS \citep{ecmwf-2019}. The cloud to rain autoconversion coefficient $k_{\mathrm{au},l}$ is also set to $7\cdot10^{-4} s^{-1}$. Autoconversion coefficients are used as main tuning parameter and are given in Part II for other resolutions. 

$q_{l0}$ is a tunable parameter for cloud liquid water and is set to 500$\cdot$10$^{-6}$ kg.kg$^{-1}$. The ice autoconversion threshold, $q_{i0}$, is temperature-dependent, following an arctangent dependency (see Eq. A.1 in \citet{lopez-2002}). The minimum threshold $q_{i0}^{min}$ has been reduced from 10$^{-7}$ kg.kg$^{-1}$ in v6.3 to 10$^{-9}$ kg.kg$^{-1}$. This lower value reduces the occurrence of small cirrus clouds with minimal water content, which may be considered numerical artifacts. Notably, the autoconversion shows a similar behavior on a log scale plot to the formulation of \citet{ryan-2000}, as used in the MESO-NH model \citep{lac-2018}.

\subsection{Total cloud cover}
The total cloud cover $C^{\mathrm{tot}}$ and the total specific cloud water content $q_{c}^{\mathrm{tot}}$ are computed for radiation calculation and diagnostics. They are expressed as functions of the stratiform and shallow convection cloud cover and specific cloud water content: 
\begin{equation}
C^{\mathrm{tot}} =  C^{\mathrm{ls}} + C^{\mathrm{sh}} - C^{\mathrm{ls}} C^{\mathrm{sh}}
\end{equation}
\begin{equation}
q_{c}^{\mathrm{tot}} = (1-C^{\mathrm{sh}})q_{c}^{\mathrm{ls}} + q_{c}^{\mathrm{sh}}
\end{equation}
Note that the total cloud amount does not include any contribution from the deep convection parameterization.

\subsection{Gravity wave drag}
\label{physics:gwd}
The orographic gravity wave drag parameterization in ARP-GEM1 is identical to that in v6.3, detailed by \citet{catry-2008} and based on \citet{lott-1997} and \citet{lott-1999}.
The vertical transport and deposition of momentum by nonorographic gravity waves is modeled using the stochastic approach described in \citet{lott-2012}. Nonorographic gravity waves are linked to convective sources through surface precipitation \citep{lott-2013}, and to frontal sources via potential vorticity \citep{de_la_camara-2015}. When used with 91 vertical levels, as in v6.3, this parameterization ensures a robust representation of the stratosphere, notably producing realistic simulations of the Quasi-Biennial Oscillation \citep{rao-2020} and a accurate depiction of extratropical variability \citep{ayarzaguena-2020}. In the low-top configuration used in ARP-GEM1 (see Section \ref{sec:accelerations}\ref{sec:levels}), the nonorographic gravity wave parameterization is disabled.

\subsection{Radiation}
\label{physics:rad}
Compared to v6.3, the radiation scheme has been updated and is now based on the ecRad radiation scheme \citep{hogan-2016}. In particular, the use of ecRad improves the representation of shortwave computations by employing the Rapid Radiative Transfer Model for GCMs \citep[RRTM-G,][]{mlawer-1997}, a correlated-k model for gas absorption, for both shortwave and longwave radiation. The use of RRTM-G for SW radiation (14 bands versus 6 bands for the previous version) notably reduces biases (not shown) when the SW radiation scheme is evaluated for clean-sky cases against line-by-line reference radiative transfer models \citep{pincus-2015}. ecRad also incorporates the Monte-Carlo Independent Column Approximation (McICA) scheme, a radiation solver that uses a stochastic generator to simulate clouds \citep{pincus-2003}. 

The handling of aerosol radiative effects is identical to that in v6.3. The ozone mixing ratio is represented as monthly-averaged climatological values. The liquid and ice cloud optical properties are computed separately in each of the 30 RRTM-G shortwave and longwave spectral bands. Liquid-droplet scattering properties depend on the effective liquid radius and are expressed in terms of a polynomial approximation computed using SOCRATES \citep{edwards-1996}. As in v6.3, the effective liquid radius depends on cloud droplet number concentration \citep{menon-2002}, in order to account for the first aerosol indirect effect. For the ice clouds, ice-particle scattering properties depend on the effective ice radius \citep{baran-2016}, which is computed as a function of ice water mixing ratio and temperature \citep{sun-2007}. 

The liquid and ice cloud optical depths are ultimately rescaled using an inhomogeneity factor for the SW radiation, both set to a value of 0.78 at 55-km resolution. Given the large uncertainties in cloud representation and cloud-radiation interactions, the inhomogeneity factors serve as a final tuning parameter that integrates cloud-radiation interaction model errors into a single, simple representation.

\subsection{Surface processes}
\label{physics:surface}
The representation of surface physical processes is handled within the SURFEX platform \citep{masson-2013}. SURFEX is called at every timestep by the physical component of the ARP-GEM1 model, with implicit treatment of the surface-atmosphere coupling. 
SURFEX uses a tile approach, considering three surface types: ocean (including sea ice), lakes, and land (including urban areas treated as rock surfaces). The surface physical processes used are similar to those in v6.3. They are documented and evaluated in \citet{decharme-2019}. The surface temperature and surface albedo over sea ice are computed using a one-dimensional version of the sea ice model GELATO version 6 \citep{voldoire-2019}.

Relative to the CMIP6 version, differences in SURFEX and GELATO arise from code adaptations and bug fixes made to address precision reduction (see Section \ref{sec:accelerations}\ref{sec:singp}), accommodate OpenMP parallelism (see Section \ref{sec:accelerations}\ref{sec:mpiopenmp}), perform computations on a coarsening grid (see Section \ref{sec:accelerations}\ref{sec:coars}), and optional choices. These choices were dictated by the desire to reduce the computational cost associated with surface processes. This is the case for the calculation of surface albedo over the oceans: use of the simple parameterization described in \citet{taylor-1996} instead of that by \citet{seferian-2018}. This is also the case for the use of only three vegetation types/patches (instead of 12) for the high-resolution ($\leq$ 25 km) versions of the ARP-GEM1 model (see Part II). Finally, the coupling with the C-TRIP model for aquifer and floodplain parameterization \citep{decharme-2019} has been disabled.

\section{Acceleration factors}
\label{sec:accelerations}

\subsection{Overall Optimizations}
\label{sec:overalloptim}
Large parts of the code have been significantly rewritten to enhance clarity, facilitate compilation and debugging, reduce memory usage, and improve computational efficiency. This was achieved by eliminating redundant or unnecessary calculations, removing useless setups and initializations, discarding superfluous arrays and/or optimizing array usage.
The number of subroutines has been reduced by a factor of 3, the number of lines of code by a factor of 5, and the compilation time by more than a factor of 10. Additionally, the memory usage has been reduced by approximately 40\%. With very few exceptions, the entire code is written in Fortran 90 and relies on only a few external libraries, thereby facilitating portability.

Parts of the dynamical core have been rewritten, in particular to allow the use of fewer three-dimensional arrays to manage the terms of the semi-implicit scheme. This development reduces the overall cost of the dynamics. Concerning the physical component, the adoption of a coarsened grid for radiation and surface computations (see Section \ref{sec:accelerations}\ref{sec:coars}) necessitated a complete restructuring of the main physics routine. This routine has been divided into three distinct parts: the first part handles calls to major physical routines; the second part manages radiation and surface computations; and the final part focuses on final tendency computations and the increment of prognostic variables. This development was also motivated by the goal of rewriting the computations of tendencies more clearly and better managing the coherence between the different variables used to describe the water mass. It results in both improved readability of the physical code and a reduction in the overall computational cost of the physics (see Section \ref{sec:results}). Lastly, in addition to the developments and simplifications presented in Section \ref{sec:physics}, most physical routines have been optimized and cleaned. For example, microphysical processes have been merged into a single routine, and unnecessary 3D variables have been removed.

\subsection{Single precision}
\label{sec:singp}

The numerical precision of real numbers is reduced from double precision (64-bit) to single precision (32-bit). This approach exploits the chaotic behavior and parameter uncertainty in numerical simulations, assuming that rounding errors remain insignificant in representing atmospheric flow \citep[e.g.,][]{palmer-2014}. Using 32-bit precision arithmetic reduces the data transferred between the CPU and memory units, as well as communications between computational nodes, thus reducing runtime. It has been successfully tested in weather and climate modeling \citep{vana-2017, nakano-2018}.

The transition from double to single precision is achieved by modifying the Fortran KIND parameter for all real numbers. Additional code modifications are necessary to support both precisions, including replacing thresholds with intrinsic precision functions, ensuring input/output files are read and written with the correct precision, and adapting operations to avoid runtime issues such as floating-point underflows and overflows. Specifically, the work has focused on these technical changes for parts of the code not addressed by the ARPEGE/IFS consortium, including some specific physical parameterizations and surface parameterizations (SURFEX and GELATO).

\subsection{Inputs/Outputs}
\label{sec:io}

\subsubsection{Prognostic and semi-prognostic variables}

The prognostic fields of the model are initially read and ultimately written in spectral space (for winds, temperature, surface pressure) or grid-point space (for specific humidity, hydrometeors and TKE) within 'historical' files. These files use the FA (\textit{Fichier ARPEGE}) format \citep{clochard-2002}. Along with the field values, these files contain metadata such as model resolution, Gaussian Grid information, and the list of vertical levels. The FA toolkit has been updated and adapted in the ARP-GEM model to accommodate the reduction in numerical precision and to handle very high resolutions.

\subsubsection{Netcdf outputs with XIOS}

For the writing of diagnostic outputs, another approach is employed. To facilitate the production of numerous diagnostics required for CMIP6, ARPEGE-Climat v6.3 (and the SURFEX platform) was interfaced with XIOS \citep{meurdesoif-2017}. XIOS is a parallel I/O server software that uses additional MPI tasks to perform postprocessing operations such as time sampling, spatial remapping, and vertical interpolation, and to write output fields during the model’s computation. The first implementation of the ARPEGE-XIOS interface was completed for ARPEGE-Climat v6.3 \citep{voldoire-2019}.

This interface has been largely updated in ARP-GEM1. In v6.3, the interface uses parts of the previous internal software for post-processing and computes spectral transformations. The new upgrade enables the direct transmission of grid-point model values to the XIOS servers, bypassing useless (and costly) transformation operations. This optimization significantly reduces diagnostic writing operations while keeping results unchanged. Moreover, in v6.3, the temporal sampling of fields was fixed to one timestep for most variables. In ARP-GEM1, this frequency is adjusted: for some variables, such as precipitation, it remains at the timestep, while for radiative fields it has been increased to one hour, and to six hours for 3D outputs. This adjustment reduces the cost of producing diagnostics while maintaining the accuracy of the results. 

\subsection{Mass conservation}
\label{sec:conserv}
 
The semi-Lagrangian scheme is inherently non-conservative \citep[e.g.,][]{williamson-1994}. In v6.3, as discussed in Section \ref{sec:dynamics}\ref{sec:SL_scheme}, a mass fixer is applied at every timestep to the hydrometeors and specific humidity. In addition, a procedure is applied to ensure the global conservation of atmospheric dry mass and total water content at each timestep. Dry mass is conserved via a redistribution using an additive correction applied to surface pressure, while total water content is preserved through a multiplicative correction to the surface precipitation flux. Without these corrections, climate simulations can experience significant mass drift after a few decades of integration. This procedure is computationally expensive, as it requires calculating global norms for surface pressure and water content, necessitating the collection of parallelized data across the entire grid. As the model resolution and the number of parallel tasks increase, the cost of this procedure becomes prohibitive.

In the ARP-GEM1 model, we do not use the mass fixer anymore, as we use a linear horizontal interpolation in the SL scheme for the hydrometeors. We also optimize the method for global conservation. We first implement and use an optimized version of the global norm computation. Second, and more importantly, we use a temporal coarsening method. While the conservation procedure is still applied at every timestep, the costly part of the fixer -- the computation of the global norms -- is executed at a lower frequency (typically every 24 timesteps). In addition, the additive correction for surface pressure is replaced by a multiplicative correction, preserving the horizontal gradient of surface pressure \citep{williamson-1994}.

Every $N$ timesteps, the global air mass $\langle p_{n} \rangle$ and the total water vapor mass $\langle q_{n} \Delta p_{n} \rangle$ are computed and a correction factor $\alpha$ is applied to keep the dry air mass equal to the initial dry air mass $M_{\mathrm{dry}}$
\begin{equation}
(1 + \alpha) \langle p_{n} \rangle - \langle q_{n} \Delta p_{n} \rangle = M_{\mathrm{dry}}
\end{equation}
At every timestep, we apply a multiplicative correction and compute the surface pressure after the mass fixer, $\tilde{p}_{n}$, as follows:
\begin{equation}
\tilde{p}_{n} = \left(1 + \frac{\alpha}{N}\right) p_{n}
\end{equation}
Note that we have performed a similar development for water mass conservation, but it has not yet been tested, as we currently do not employ water correction in AMIP experiments.

This new procedure allows for small corrections (on the order of a few percent) while significantly reducing the model's computational cost. Given that global tracer-mass conservation errors generally decrease with increasing resolution, the temporally coarsened mass fixer appears to be a suitable approach.

\subsection{Vertical levels}
\label{sec:levels}

\begin{figure}[h]
\centerline{\includegraphics[width=19pc]{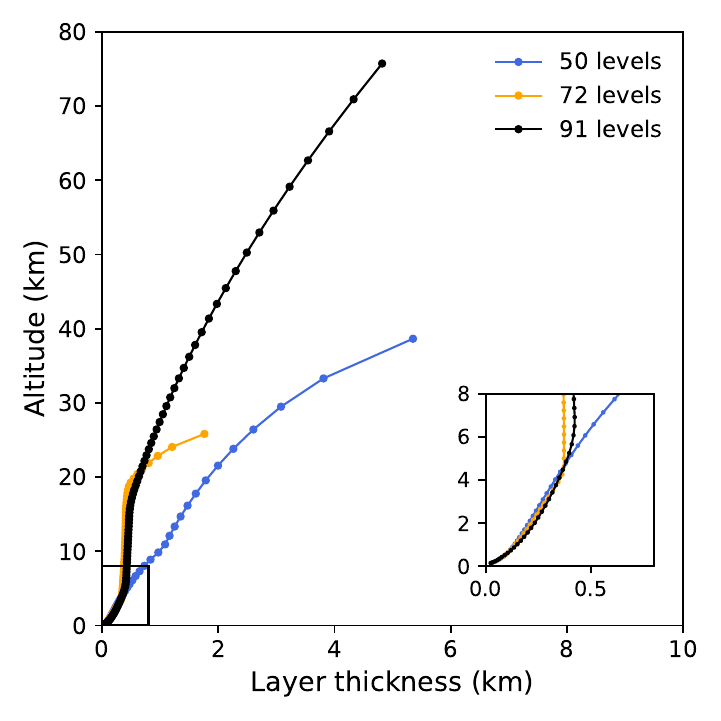}}
\caption{Vertical resolution used in ARPEGE‐Climat v6.3 (91 levels, black) and in ARP-GEM1 (50 levels, blue). The 72-level version is also plotted (orange). The y-axis refers to the log-altitude (km) defined as $z = 7 \ln(p_0/p)$, where $p_0$ = 1013 hPa and $p$ is the pressure at interface layers. A zoom over the first 8 km is shown in the bottom right corner.}
\label{fig:levs}
\end{figure}

The vertical discretization in the ARPEGE model follows a terrain-following pressure hybrid coordinate system \citep{simmons-1981}. ARPEGE-Climat v6.3 employs 91 vertical levels, with 33 layers in the stratosphere and 8 layers in the mesosphere, as illustrated in Figure \ref{fig:levs}. 

In contrast, the ARP-GEM1 model uses a low-top configuration with a reduced set of 50 vertical levels, focusing mainly on the representation of the troposphere. To set the hybrid pressure coordinate, the method described in \citet{benard-2008} has been slightly adapted. This configuration features fewer levels in the upper troposphere and a significantly reduced number of layers in the stratosphere, resulting in a model top at approximately 40 km. Despite the reduction in vertical layers, the boundary layer discretization remains similar to that of the 91-level configuration. Note that a 72-level discretization has also been developed and a sensitivity to the vertical resolution is shown in Part II.

\subsection{Grids}
\label{sec:cubic}

\subsubsection{Cubic spectral grid}
Winds, temperature, the logarithm of surface pressure, and orography are represented by a truncated series of spherical harmonics. In the spectral space, the horizontal resolution is thus expressed by the cut-off spectral truncation number $n$ of the spherical harmonics series expansion. 

The distribution of points allowing exact transformation from grid-point space to spectral space is a Gaussian grid, defined by $N_{g}$ Gaussian latitudes along a meridian, with at least $N_{g} \geq (2n+1) / 2$ Gaussian latitudes and $2n + 1$ longitudes along a given Gaussian latitude. The horizontal resolution in grid-point space is simply the distance between two Gaussian latitudes. A grid with at least $2n + 1$ longitudes is called a \textit{linear} grid as it differs from a \textit{quadratic} grid (at least $3n + 1$ longitudes), that causes no aliasing to occur in the quadratic terms, or a \textit{cubic} grid (at least $4n + 1$ longitudes), avoiding aliasing in the terms involving triple products \citep{wedi-2014}. 

ARPEGE-Climat v6.3 uses a linear grid. Linear grids were historically chosen because they minimize the number of grid points for a given resolution \citep{cote-1988} and because semi-Lagrangian advection partially mitigates nonlinear effects. In ARP-GEM1, we use a cubic grid, enhancing efficiency of the spectral method while maintaining accuracy \citep{wedi-2014}.

\subsubsection{Octahedral reduced Gaussian grid}

In v6.3, the number of longitudes is reduced towards the poles to keep relative distances between points approximately constant \citep{hortal-1991}. In ARP-GEM-1, this reduced Gaussian grid has been replaced by an octahedral reduced Gaussian grid \citep{malardel-2016}, which optimizes the total number of points around the globe by introducing a regular reduction of points per latitude circle, allowing for a quasi-uniform resolution. This new reduced grid contains about 25\% fewer points than the original reduced Gaussian grid.

In the 55-km resolution configuration of ARP-GEM1, we use a TCo179 grid (where T stands for truncation, C for cubic, and o for octahedral), corresponding to a spectral truncation of $n=179$ and an octahedral reduced grid with $N_{g} = 360$ Gaussian latitudes and $720$ longitudes along equatorial Gaussian latitudes. The reduced Gaussian grid contains 136,000 points.

\subsection{Spatial coarsening for radiation and surface processes}
\label{sec:coars}

\begin{table*}
\caption{Elapsed time and computational cost of the ARPEGE-Climat v6.3 and the ARP-GEM1 model. Elapsed time (unit : sec) refers to the total model runtime. CHSY refers to elapsed time multiplied by the number of cores and is expressed in kh.cores. The runtime are estimated from a 2-year simulation, at 55 km run on 9 nodes (= 1152 cores).}
\label{tab2}
\begin{center}
\begin{tabular}{lrr}
\topline
model & elapsed time (sec) & CHSY (kh.cores) \\
\midline
ARPEGE-Climat v6.3 (55 km) & 17180 & 5.50 \\
ARP-GEM1 (55 km) & 1120 & 0.35  \\
\botline
\end{tabular}
\end{center}
\end{table*}

The radiative transfer model often accounts for a significant portion of the model's runtime. Several strategies can be employed to reduce its computational cost while preserving accuracy \citep{pincus-2019}. A commonly used method involves using a radiative timestep -- i.e., the interval at which the complete radiative transfer model is called -- that is larger than the model timestep. Another strategy consists of performing radiation calculations on a grid with a coarser resolution than the model grid \citep{morcrette-2008}.

In the ARP-GEM1 model, we combine both approaches and also use spatial coarsening for surface fluxes. At a 55 km resolution, we use a 2-hour radiative timestep (one call every eight model timesteps). The computation of atmospheric heating rate profiles between consecutive radiation calls has been modified compared to v6.3. At each model timestep, shortwave profiles are updated with the change in solar zenith angle and corrections are now made to account for the modified path length of solar radiation \citep{manners-2009}. Longwave profiles are updated to account for changes in local skin temperature \citep{hogan-2015}.

Radiation and surface calculations are performed on a grid with a coarser resolution than the model grid. Interpolation between model and coarsened grids is performed using existing interfaces within the code for the semi-Lagrangian scheme, with an effort made to minimize the number of variables involved. At 55 km, we use a spatial grid coarsening factor (ratio of original to coarsened grid resolution) of about 2.8 for both radiation and surface. Different values are applied at higher resolutions (see Part II). Note that at LES resolution, low-level stratocumulus clouds are, to a first approximation,  not highly sensitive to spatial averaging of radiation \citep{bellon-2016}. Applying the same coarsening factor to surface fluxes ensures consistency with surface radiation. However, the topographic resolution is retained as in the core model, preserving the benefits of higher resolution for accurately representing dynamical effects related to topography.

Finally, the use of a coarsened grid benefits also to the writing of surface and radiation outputs since the  variables are exchanged and written through the XIOS server at the coarsened resolution.

\subsection{MPI and OpenMP parallelizations}
\label{sec:mpiopenmp}

The ARPEGE/IFS code uses both distributed and shared memory parallelization techniques. Distributed memory parallelization, where each processor/task manages only a subsection of the total data, is achieved through Message Passing Interface \citep[MPI,][]{gropp-1999}, with data distribution managed by a high-level and efficient transposition algorithm \citep{barros-1995}.

Shared memory parallelization, which manages on-node computations, is achieved through threading with OpenMP. In v6.3, the hybrid MPI/OpenMP parallelization was not functional, particularly due to limitations in the SURFEX component. We have addressed these limitations in the ARP-GEM1 model, making OpenMP parallelization effective. In the results presented here and in Part II, only MPI parallelization is active. However, the hybrid MPI/OpenMP parallelization will be used in future high-resolution simulations to reduce memory usage.

\begin{figure*}[h]
\centerline{\includegraphics[width=33pc]{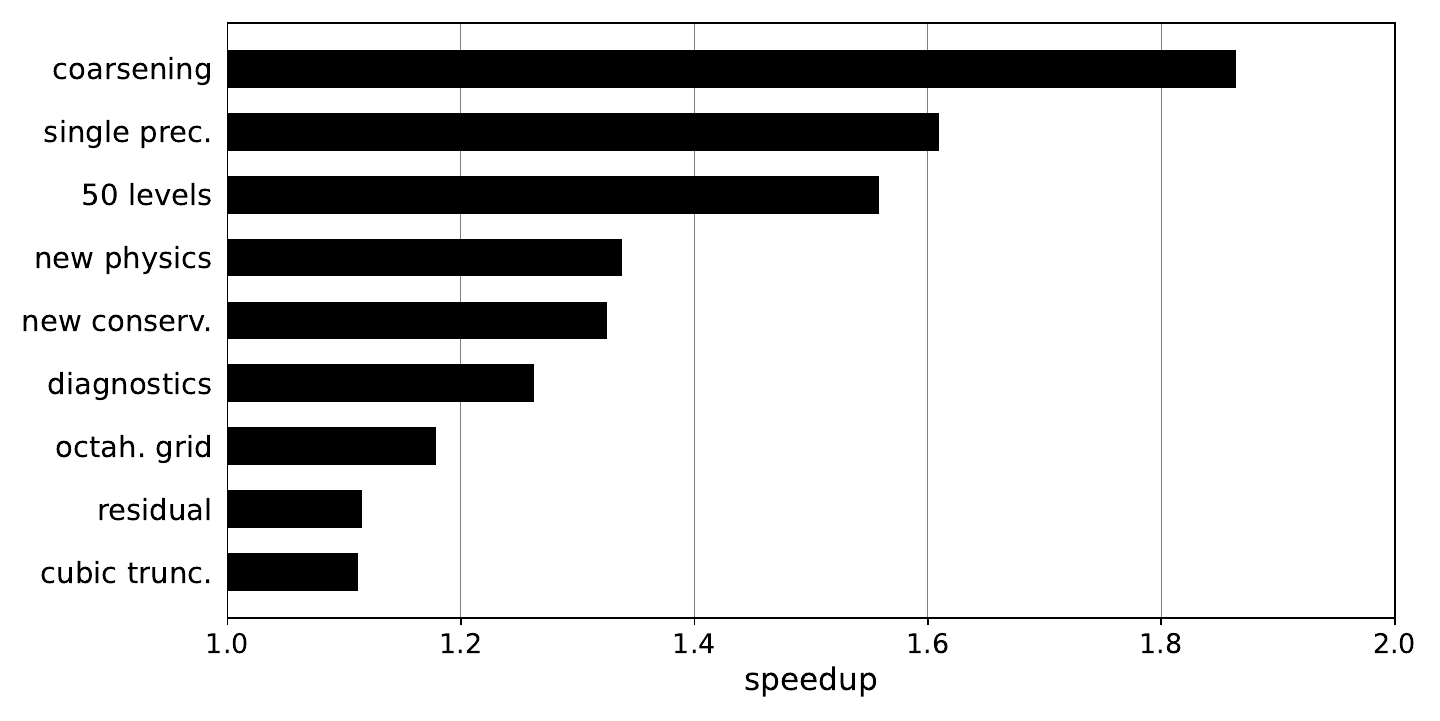}}
\caption{Estimated speedup of each acceleration factor: coarsened grid for radiation and surface processes (\textsf{coarsening}, Section \ref{sec:accelerations}\ref{sec:coars}), single precision (\textsf{single prec.}, Section \ref{sec:accelerations}\ref{sec:singp}), reduced set of 50 vertical layers (\textsf{50 levels}, Section \ref{sec:accelerations}\ref{sec:levels}), enhanced and optimized physical parameterizations (\textsf{new physics}, Sections \ref{sec:physics} and \ref{sec:accelerations}\ref{sec:overalloptim}), optimized mass conservation method (\textsf{new conserv.}, Section \ref{sec:accelerations}\ref{sec:conserv}), optimized interface with XIOS server (\textsf{diagnostics}, Section \ref{sec:accelerations}\ref{sec:io}), octahedral reduced Gaussian grid (\textsf{octah. grid}, Section \ref{sec:accelerations}\ref{sec:cubic}), spectral cubic grid (\textsf{cubic trunc.}, Section \ref{sec:accelerations}\ref{sec:cubic}), residual, mainly resulting from code optimizations, particularly in the dynamical core (\textsf{Residual}, Sections \ref{sec:dynamics} and \ref{sec:accelerations}\ref{sec:overalloptim}).}
\label{fig:speedup}
\end{figure*}

\section{Speed-up results}
\label{sec:results}

\subsection{Overall gain in computational cost}

This section focuses on computational performance results. An evaluation of the model is provided in Part II. The computational performance of a climate model is commonly measured in core hours per simulated year (CHSY), which is the product of the model runtime for one simulated year and the number of cores allocated. This value may depend on the hardware specifications. Another frequently used metric, simulated years per day (SYPD), can be derived from CHSY and the number of cores allocated. ARPEGE-Climat v6.3 and ARP-GEM1 runs on the Météo-France supercomputer Belenos\footnote{https://www.top500.org/system/179853}. Each computational node is equipped with two AMD Epyc Rome processors, each with 64 cores operating at at 2.25 GHz.  

Table \ref{tab2} summarizes the computational cost of ARP-GEM1 (with a horizontal resolution of 55 km) and compares it to that of ARPEGE-Climat v6.3 with the same grid-point horizontal resolution. The computational cost of v6.3 is likely to be relatively standard: the cost of the intermediate version of the CNRM climate model analysed in \citet{balaji-2017} is a median value (their Table 1).  At equivalent intermediate horizontal resolutions and using the same hardware platform, the ARP-GEM1 model is 15 times faster than the atmospheric component of CNRM-CM6-1-HR. With a thousand cores, the SYPD reaches a value of approximately 77. Note that this acceleration is achieved by keeping a 900 s timestep, which can be increased to 1200 s at 55 km resolution.

\subsection{Contribution of each acceleration factor}

In this section, we provide an estimation of the contribution of each acceleration factor to the overall performance. The specific gains from each optimization depend on the model's initial design and configuration; thus, the values provided serve as an indication of the cost associated with each optimization. The relative contributions of each optimization are summarized in Figure \ref{fig:speedup}.

When possible, speedups are estimated by comparing the elapsed time between two paired experiments: one using the final ARP-GEM1 code and the other with the corresponding speedup factor disabled. For example, the impact of precision reduction is evaluated by comparing an ARP-GEM1 simulation run in double precision with one run in single precision. This method is employed to assess the effects of radiation and surface coarsening, reduced vertical levels, the use of single precision, the octahedral grid, and cubic truncation.

To estimate the optimization of the diagnostics due to changes in the XIOS interface with ARP-GEM1, paired experiments (with and without a set of standard XIOS outputs) are carried out with ARPEGE-Climat v6.3 and ARP-GEM1. The value of the corresponding speedup (labeled as diagnostics in Figure \ref{fig:speedup}) is the ratio between the cost of outputing standard netcdf diagnoses (difference between a simulation with and without XIOS) between the v6.3 version and the ARP-GEM1 model. 

The remaining factors (new physics, new mass conservation method) are estimated indirectly by comparing simulations with v6.3 and ARP-GEM1 at equivalent grid-point resolution, precision (ARP-GEM1 is run in double precision), and number of vertical levels. The overall costs of the physics and of the conservation procedure are estimated with the use of the analytical tool, Dr. Hook \citep{saarinen-2005}, which gathers performance profiling data for each subroutine.

\begin{figure*}[h]
\centerline{\includegraphics[width=33pc]{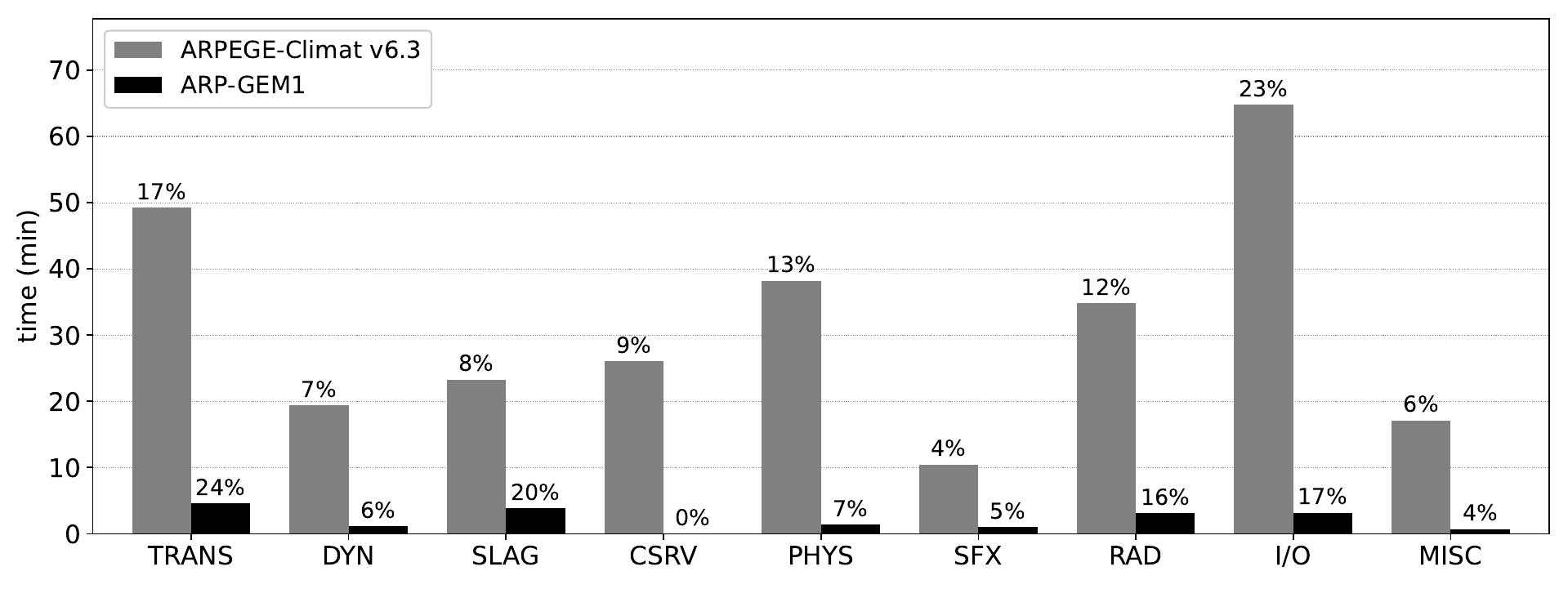}}
\caption{Decomposition of the overall computational cost of ARPEGE-Climat v6.3 (gray) and ARP-GEM1 (black) into their main components, as described in the text. The overall time (in minutes) represents the runtime for one simulated year at a resolution of $55$ km.}
\label{fig:dhkall}
\end{figure*}

The overall speedup factor of 15 results from the combined use of the multiple acceleration factors rather than a single dominant one, with individual speedup factors ranging from 1.1 to 1.8. The residual speedup (close to 1.1) is calculated by dividing the overall speedup by the product of all estimated speedups. It results mainly from various code optimizations notably in the dynamical core and initial setups but also includes effects from interactions between optimizations and uncertainties in the estimates.

The most significant speedup, approximately 1.8, is achieved using a coarsened grid for radiative and surface computations. This depends on the grid coarsening factor. In this case, the coarsening factor of 2.8 reduces the number of columns viewed by the radiation and surface schemes by about eight. Bit reduction for floating-point representation results in a speedup of nearly 1.6, consistent with reported values in the literature. The speedup from reducing the number of vertical levels from 91 to 50 is almost 1.6, which is lower than the ratio of 91 to 50. This is because some parts of the code (e.g., surface processes) do not depend on the number of levels, and certain physical parameterizations are only active in specific atmospheric layers.

The use of an octahedral grid reduces the number of grid points by approximately 25\%, though this reduction benefits only computations directly involving grid points. Conversely, employing a cubic grid, with a truncation number halved compared to a linear grid, decreases spectral computations. When combined together, the cubic truncation and the icosahedral reduced grid configuration achieves a speedup of approximately 1.3.

The other speedups are specific to the ARP-GEM model. Improvements in interfacing the I/O server XIOS results in a speedup of almost 1.3. This speedup can vary depending on the number of diagnostics generated by the model. Optimizing the conservation step with the use of temporal coarsening leads to a speedup of approximately 1.3. Notably, the previous conservation method's cost increased significantly with higher resolution, an issue that the new method effectively addresses. 

Finally, enhancements, including optimizations and simplifications to the physical parameterization schemes and the main code achieve an additional speedup factor of 1.4. We note that the use of the ecRad scheme introduces a marginal additional cost due to the increased complexity of the SW scheme. This additional cost is relatively offset by the reduced frequency of radiation calculations compared to the initial model and is absorbed by the spatial coarsening implemented in the new model.

\subsection{Decomposition per model component}

\subsubsection{General}

\begin{figure*}[h]
\centerline{\includegraphics[width=33pc]{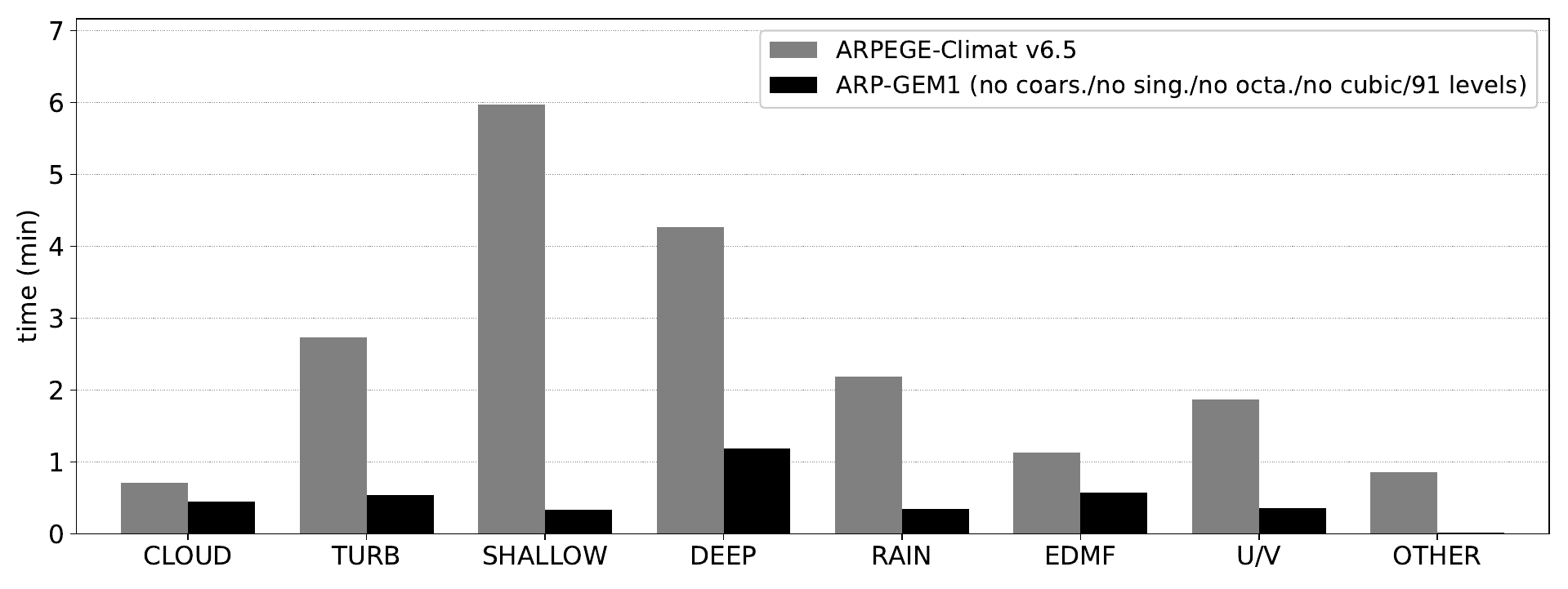}}
\caption{Distribution of computational cost of physical parameterizations in ARPEGE-Climat v6.5 and ARP-GEM1. Time (in minutes) represents the runtime for one simulated year.}
\label{fig:dhkphys}
\end{figure*}

Figure \ref{fig:dhkall} compares the computational cost distribution between v6.3 and ARP-GEM1. The cost of the dynamical core includes the transpositions between grid-point and spectral spaces (TRANS), the semi-implicit spectral computations (DYN), and the semi-Lagrangian grid-point computations (SLAG). The CSRV category refers to the cost of the subroutines related to the mass conservation procedure. The cost of physical parameterizations is divided into physics (PHYS), surface processes (SFX), and radiation (RAD). The I/O category encompasses the writing and reading of historical files, as well as the preparation and writing of diagnostics. Finally, the MISC category primarily includes initial setups.

Due to the combination of acceleration factors, all of the costs have been significantly reduced in ARP-GEM1. The cost of the dynamical calculations is reduced thanks to the use of cubic truncation (TRANS, DYN), the use of octahedral grid (SLAG, TRANS), the use of reduced precision and fewer vertical levels (SLAG, TRANS and DYN). The dynamics may also benefit from optimizations in other parts of the code by reducing imbalance and communication delays. The computation of the terms used by the semi-implicit scheme has also been optimized (DYN).

Likewise, the cost of the new conservation procedure (CSRV) is close to zero whereas the mass fixer accounts for almost 10\% of the model cost in v6.3. This reduction is primarily due to the use of temporal coarsening techniques for the calculation of the global norms (section \ref{sec:accelerations}\ref{sec:conserv}). The cost of the physics computations (PHYS) is notably reduced and is also related to single precision, reduced vertical levels, use of octahedral grid and physics optimizations, with further details provided in the next section. The radiation (RAD) and surface (SFX) computations benefits both from the coarsened grid and from the use of single precision. The reduction of the cost of the I/O part mainly comes from the optimization of the interfacing with XIOS (as described in Section \ref{sec:accelerations}\ref{sec:io}). Finally, common parts of the code (MISC) benefit from optimizations such as removing unnecessary operations and accelerating initial setups.

Not only have all the costs been reduced, but the relative contributions have also shifted. The only relative increases are in the dynamical components, mainly SLAG and also TRANS. Excluding the mass conservation procedure, the dynamical core (TRANS + DYN + SLAG) now accounts for 50\% of the overall cost, up from 33\% in the previous model, with fluid dynamics representing the primary computational expense of the atmospheric model. This shift aligns with the original optimization goal: reducing the cost of non-dynamical components so that dynamics, as the core process of the model, becomes the dominant contributor to the overall cost.

Non-dynamical components, including physical parameterizations, radiation, and surface processes, have experienced a relative decrease in cost. Despite these reductions, some processes remain costly, leaving room for further optimizations. Current developments, such as increasing coarsening factors, would help reduce these costs, especially at higher resolutions. However, new developments are necessary for additional reductions. For instance, surface processes (SFX) remain relatively expensive and would benefit from further updates and simplifications of the surface model. 

\subsubsection{Physical parameterizations}
\label{sec:results_phys}

About 45\% of the reduction in computational cost associated with the modifications of the physics is due to a significant reduction in general aspects such as initialization, computation of tendencies, and computation of derived diagnostics (not shown). In ARP-GEM1, the cost of these side computations is almost reduced to zero (not shown). The remaining reduction in computational cost is associated with a decrease in the cost of the model's main physical schemes. Figure \ref{fig:dhkphys} compares the computational cost distribution of the physical parameterizations between ARPEGE-Climat v6.5 and ARP-GEM1. This comparison, focused on optimizations of the schemes, is conducted at equivalent grid-point resolution, precision, and number of vertical levels.

Note that the cost of ARP-GEM1 is compared to v6.5 rather than v6.3. Version 6.5 has similar routines to ARP-GEM1 (Table \ref{tab:phys}), with a classic decomposition of processes (e.g., shallow convection, deep convection, and a separate cloud scheme) and state-of-the-art convection schemes. The main difference between v6.5 and v6.3 is the cloud scheme, which is part of the turbulence scheme in v6.3 but has a similar cost, and the convection schemes. The convection scheme used in v6.3 \citep{piriou-2007, gueremy-2011} is costly due to its poor optimization and the use of a large number of prognostic variables. Its overall cost is comparable to that of the deep and (expensive) shallow convection schemes in v6.5, making the physical computations in v6.3 and v6.5 roughly equivalent.

The cost of all physical schemes in ARP-GEM1 has been notably reduced in association with the developments described in Sections \ref{sec:physics} and \ref{sec:accelerations}\ref{sec:overalloptim}. This reduction is particularly significant for the shallow convection scheme, where improvements have been made through the use of a more compact routine and more efficient code. Additionally, the cost of the deep convection scheme has been reduced due to the removal of shallow (and middle) convection in the Tiedtke-Bechtold code and modifications in the triggering mechanism.
In the case of the turbulence scheme, cost reductions have been achieved through optimizations in the mixing length computation, simplifications, and the removal of redundant computations in the main turbulence routine.

Microphysics, in particular, cloud scheme and numerical resolution of turbulent and shallow convective mixing (EDMF) have benefited from code optimizations, including reduced memory utilization and the removal of unnecessary 3D variables.
The reduction in costs for computing subscale momemtum fluxes (U/V) is related to the disabling of the non-orographic gravity wave scheme. 
Finally, the remaining part of the physical parameterization codes (OTHER) has benefited from the removal of redundant processes or their merging into other routines.

\section{Conclusion}

The development of ARP-GEM, a Global, Efficient and Multiscale atmospheric climate model, was motivated by the goal of creating an atmospheric model capable of efficiently running climate simulations at resolutions ranging from synoptic to km-scale. It aims to achieve this using a minimal set of physical parameterizations necessary for accurate climate modeling, thereby addressing the complexity that has accumulated over the years in current atmospheric models. This increase in complexity necessitates substantial cleaning and modernization, as highlighted in recent projects such as the Energy Exascale Earth System Model project \citep{caldwell-2021}.

ARP-GEM1 is based on ARPEGE-Climat v6.3 and incorporates a series of efficiency-oriented improvements (Section \ref{sec:accelerations}) alongside an enhanced suite of state-of-the-art physical parameterizations, including new shallow convection, deep convection, large-scale cloud, radiation and turbulence (Section \ref{sec:physics}).
The code has been extensively rewritten to improve efficiency, clarity, facilitate debugging, and support future development. These changes enhance code maintenance, ease adaptation for new developers, and improve portability, making it easier to adapt the code for new hardware platforms. 

The ARP-GEM1 model is approximately 15 times faster than ARPEGE-Climat v6.3 while maintaining accurate simulations and even improving model performance (see Part II). This significant acceleration is attributed to a combination of several optimizations rather than a single dominant factor (Section \ref{sec:results}). These acceleration factors include optimizations in the general code, dynamics, and physics, the use of coarsened grids for radiative and surface computations, bit reduction for floating-point operations, adoption of octahedral and cubic grids, improvements in I/O operations, optimization of the mass conservation step, and refinements to physical parameterizations. While some developments are specific to ARP-GEM, the results presented here provide insights into various possibilities for efficiency-oriented improvements.

In ARP-GEM1, the contribution of dynamics to the overall computational cost has risen to approximately 50\%, reflecting its central role in the model's performance. This increase aligns with the optimization goal of making dynamics the primary cost driver. 
In the present version, some components of the model remain relatively costly, such as radiation, surface processes, and I/O operations, necessitating further optimization.
This is especially true at resolutions high enough to require a non-hydrostatic core, which is more computationally expensive. 

Recent optimizations of the ecRad module could be considered \citep{ukkonen-2024}.
The importance of surface processes could be reduced by decreasing the surface model's complexity, without significantly impacting model results. The atmospheric model would also benefit from improved I/O efficiency, which is a major contributor to model cost, especially in simulations with substantial output demands. 
At high resolutions and smaller timesteps, temporal and spatial coarsening factors (radiation, surface) may be increased, and the cost of subgrid parameterizations should decrease, given that more processes are explicitly represented, beginning with the elimination of deep convection. Parameterizations may be simplified, with the potential exception of microphysics, which may benefit from a more complex representation (e.g., more species) to account for the increased realism associated with high resolution.

Finally, the code is currently optimized for large CPU supercomputers but needs to become hardware-agnostic in the future, possibly through the use of Domain-Specific Languages. The acceleration approach presented here could be combined with architectural changes. The results suggest that with such additional acceleration, global kilometer-scale climate simulations should become feasible.

\acknowledgments

\datastatement
For the atmospheric part of the ARP-GEM1 code and model outputs, please contact the authors. The SURFEX code is available under a CECILL‐C License at the SURFEX website (http://www.umrcnrm.fr/surfex). XIOS can be downloaded from the XIOS website (https://forge.ipsl.jussieu.fr/ioserver).


\bibliographystyle{ametsocV6}

\end{document}